\documentclass[prl,twocolumn,showpacs]{revtex4}

\usepackage{epsfig}

\begin{document}
\newcommand{\ltwid}{\mathrel{\raise.3ex\hbox{$<$\kern-.75em\lower1ex\hbox{$\sim$}}}}
\newcommand{\gtwid}{\mathrel{\raise.3ex\hbox{$>$\kern-.75em\lower1ex\hbox{$\sim$}}}}
\newcommand{\bra}{\langle}
\newcommand{\ket}{\rangle}

\title{Real time evolution using the density matrix renormalization group}

\author{Steven R.~White and Adrian E. Feiguin}

\affiliation{Department of Physics and Astronomy\\
University of California, Irvine, CA 92697}

\date{\today}

\begin{abstract}

We describe an extension to the density matrix renormalization group method 
incorporating real time evolution into the algorithm. Its application to 
transport problems in systems out of equilibrium and frequency 
dependent correlation functions is discussed and illustrated in several 
examples. We simulate a scattering process in a spin chain which generates
a spatially non-local entangled wavefunction. 

\end{abstract}
\pacs{PACS }

\maketitle


The density matrix renormalization group (DMRG) \cite{dmrg} is perhaps the 
most powerful method for simulating one dimensional quantum lattice 
systems.
DMRG was originally formulated as a ground state method. Later, it
was generalized to give frequency dependent spectral functions 
\cite{karen,kuehner}. The best spectral method, Jeckelmann's dynamical 
DMRG \cite{jeckelmann}, yields extremely 
accurate spectra. However, it is limited to only one momentum and one 
narrow frequency range at a time. Constructing an entire spectrum for a 
reasonable grid in momentum and frequency space can involve hundreds of 
runs. 

An alternative approach to dynamics with DMRG is via a real time 
simulation.
Cazalilla and Marston introduced a real-time DMRG and used
it to calculate the time evolution of one dimensional systems under an
applied bias\cite{marston}.
In their approach, the DMRG algorithm is only used to calculate the ground
state, and the time evolution in obtained by integrating the
time-dependent Schr\"{o}dinger equation in a fixed basis.
Consequently, one expects it to lose accuracy when the wavefunction
starts to differ significantly from the ground state.
In the systems studied by Cazalilla and Marston, the time evolution
could be carried out for a reasonable length of time before this happened.
Luo, Xiang, and Wang\cite{comment} showed how to
construct a basis which applies to a time-evolving
wavefunction over a whole range of times simultaneously. This approach was 
shown to be more accurate than that of Cazalilla and Marston,
but it is not very efficient---the basis must be quite
large to apply to a long interval of time, and the whole time evolution is
performed at every DMRG step.

Recently, Vidal developed a novel time-dependent simulation 
method for near-neighbor one dimensional systems which 
overlaps strongly with DMRG \cite{vidal}.  The crucial
new idea of the method is to use the Suzuki-Trotter decomposition for
a small time evolution operator $\exp(-i\tau H)$. 
The second order Suzuki-Trotter break-up is
\begin{equation}
e^{-i\tau H} \approx 
e^{-i\tau H_A/2} 
e^{-i\tau H_B} 
e^{-i\tau H_A/2},
\label{simpletrotter}
\end{equation}
where $H_A$ contains the terms of the Hamiltonian for the even links, and 
$H_B$ for the odd. The individual
link-terms $\exp(-i\tau H_j)$ (coupling sites $j$ and $j+1$) 
within $H_A$ or $H_B$ commute. Writing the wave function in matrix
product form (which underlies the DMRG block form\cite{rommer}), Vidal
showed that one can apply each link-term directly to the wavefunction, 
exactly and efficiently.  After each such application,
a Schmidt decomposition, equivalent to diagonalizing the DMRG density
matrix, is performed to return the wavefunction to the matrix product form.
One applies all the $H_A$ terms, and then all the $H_B$ terms, etc.

Although this method seems very efficient, a number of aspects are novel
for DMRG users, stemming from the fact that one does not ordinarily 
deal with the matrix product representation directly. Implementing this
idea into a DMRG algorithm may be time consuming and may require a very 
substantial rewriting of one's program.

In this paper we take the key idea of the Suzuki-Trotter decomposition,
but we modify it and apply it in a more natural way within the
context of DMRG. The result is a surprisingly simple yet very powerful
modification of the algorithm for real-time dynamics which we believe can 
be incorporated into a typical program in only a day or two of 
programming. We illustrate
the approach with real-time simulations which set a new paradigm
for the size and accuracy obtainable.

The standard DMRG representation of the wavefunction at a particular step
$j$ during a finite system sweep is
\begin{equation}
|\psi\ket = \sum_{l\alpha\beta r} \psi_{l\alpha_j\alpha_{j+1} r} 
|l\ket |\alpha_j\ket |\alpha_{j+1}\ket |r\ket.
\label{dmrgwfn}
\end{equation}
Here we have a left block containing many sites (with states $l$), 
two center sites (with states $\alpha_j$, $\alpha_{j+1}$), and a right 
block (states $r$).
The states $l$ and $r$ are formed as eigenvectors of a density matrix,
and represent a highly truncated but extremely efficient basis for representing
the ground state, plus any other targeted states which have been included
in the density matrix. Now suppose we have an arbitrary operator $A$ acting 
only on sites $j$ and $j+1$. This operator
can be applied to $|\psi\ket$ exactly, and reexpressed in terms of the same 
optimal bases, with
\begin{equation}
[A\psi]_{l\alpha_j\alpha_{j+1} r} 
= \sum_{{\alpha_j}'{\alpha_{j+1}}'} 
A_{{\alpha_j}{\alpha_{j+1}};{\alpha_j}'{\alpha_{j+1}}'}
\psi_{l{\alpha_j}'{\alpha_{j+1}}'r} .
\label{applyA}
\end{equation}
If $A$ included terms for other sites, we could not write this simple
exact relation; new bases would need to be adapted to describe both
$|\psi\ket$ and $A|\psi\ket$, requiring perhaps several finite-system sweeps 
through the lattice.

This implies that we can apply the link time evolution operator 
$\exp(-i\tau H_j)$ exactly on DMRG step $j$, but at any other DMRG
step the application is approximate. Accordingly, we adapt
the Suzuki-Trotter decomposition to match the DMRG finite-system sweeps,
so that each term can be applied exactly. We decompose the time propagator as
\begin{equation}
e^{-i\tau H} \approx 
e^{-i\tau H_1/2} 
e^{-i\tau H_2/2} 
\ldots
e^{-i\tau H_2/2}
e^{-i\tau H_1/2}.
\label{mytrotter}
\end{equation}
This decomposition is good to the same order (errors of order $\tau^3$)
as the usual odd/even link decomposition. When applied $1/\tau$ times
to evolve one unit of time, the errors are $\tau^2$. The main idea is then
to apply $\exp(-i\tau H_1/2)$ at DMRG step 1, then $\exp(-i\tau H_2/2)$
at step 2, etc., forming the usual left-to-right sweep, then reverse,
applying all the reverse order terms in the right-to-left sweep.

This procedure requires one to use the step-to-step wavefunction 
transformation first developed to provide a good guess for the
Lanczos or Davidson diagonalization\cite{whitecavo}, where it can
improve run times by an order of magnitude.
It transforms the wavefunction from the basis of step $j\pm 1$
to that of step $j$.
Assuming this transformation is present in a ground state DMRG program, 
the real-time algorithm introduces only a very minor modification:
at step $j$, instead of
using the Davidson method, one evolves the transformed wavefunction 
by applying $\exp(-i\tau H_j/2)$.
Before the time evolution starts, we typically use ordinary DMRG to find
the ground state. Next, we either (1) change the Hamiltonian, or (2) apply 
an operator
to the ground state to study a new wavefunction which is a combination of 
excited states.

\begin{figure}
\begin{center}
\epsfig{file=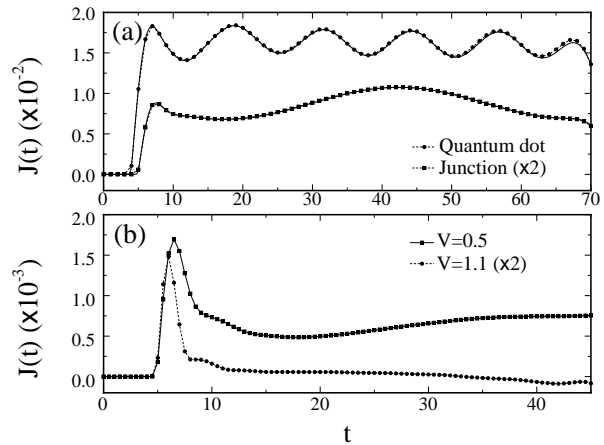,angle=-90,width=8cm}
\caption{(a) Tunneling current through a non-interacting quantum dot 
and a junction as defined in  Eqs. (2) and (5) in Ref.\cite{marston}, 
respectively. The full lines correspond to exact results. (b) Tunneling 
current through an interacting junction, with $V=0.5$ and $V=1.1$. All 
the DMRG results where obtained with $M=128$ and a time step $\tau=0.2$.}
\label{current}
\end{center}
\end{figure}

As a first test case, of type (1), 
we study the models of Eqs.(2) and (5) in Ref.
\cite{marston} corresponding to a quantum dot connected to two
non-interacting leads, and a junction between two Luttinger liquids,
respectively, driven out of equilibrium by a voltage bias.
In these cases at $t=0$ a bias in the chemical potential is turned on
as a smoothed step function, making the new Hamiltonian time-dependent. 
At each time step, the expectation value of the current operator (defined 
by Eq.(4) of Ref.\cite{marston}) is calculated. In Fig.\ref{current} we 
show the results for a chain
of length $L=64$ and the same set of parameters used in Ref.\cite{marston},
keeping only $m=128$ states and using a time step
$\tau = 0.2$. It should be compared to Figs. 1 and 2 in Ref.
\cite{marston} and Figs. 1 and 2 in Ref.\cite{comment}.
Our results exceed the accuracy obtained by the previous methods,
with only a fraction of the states.
For the non-interacting problem, the
agreement with the exact results is excellent up to times $t\sim 70$.
The main reason for obtaining higher accuracy for fixed $m$ compared to
Ref.\cite{comment} is that at any step we only need to target one 
state at one instant of time, not a whole range of times.

In method (2), after finding the ground state $| \phi \ket$, we apply an operator $A$
at $t=0$, to obtain $| \psi(t=0) \ket$, and evolve in time. We first use 
this approach to calculate time-dependent correlation functions.
In this case, we time evolve both $| \phi(t) \ket$ and $| \psi(t) \ket$,
including both as target states for the DMRG density matrix.
Although the time dependence $\exp(-iE_Gt)$ of
$|\phi(t)\ket$ is known ($E_G$ is the ground state energy), 
by evolving it we keep its representation in
the current basis. In addition, we expect a significant cancelation in the
errors due to the Suzuki-Trotter decomposition in constructing the
correlation functions.
A typical correlation function is calculated as
\begin{equation}
<\phi|B(t) A(0)|\phi> = <\phi(t)|B|\psi(t)>,
\label{greens}
\end{equation}
We use a complete half-sweep to apply $A$ to $| \phi \ket$.
In particular, if $A$ is a sum of
terms $A_j$ over a number of sites, then we apply an $A_j$ only when $j$ 
is one of the two central, untruncated sites. Thus the basis is automatically
suitable for $A_j | \phi \ket$.  During this buildup of $A$ at step $j$ we 
target both
the ground state $|\phi\ket$ and $\sum_{j'=1}^j A_{j'}|\phi\ket$. 


As an example we consider the spin-1 Heisenberg chain, with 
Hamiltonian
\begin{equation}
H = \sum_j \vec{S}_j \vec{S}_{j+1} \;,
\end{equation}
where we have set the exchange coupling $J$ to unity. This system has
a gap (the ``Haldane gap'') of $\Delta_H = 0.4105$ 
to the lowest excitations, which are spin-1 magnons at
momentum $\pi$, and a finite correlation length of $\xi = 6.03$.\cite{huse} 
The single-magnon dispersion relation has been calculated
with excellent accuracy\cite{kuehner}. However, determination of
the full magnon line is quite tedious with existing DMRG methods. Here
we demonstrate how to calculate the entire magnon spectrum with only one
time dependent DMRG run. 

\begin{figure}
\begin{center}
\epsfig{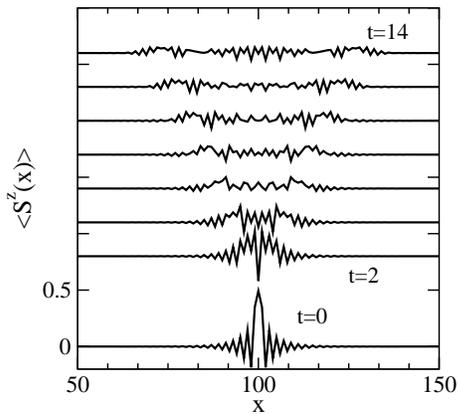}
\caption{ Time evolution of the local magnetization $\bra S^z(x) \ket$ 
of a 200 site spin-1 Heisenberg chain after $S^+(100)$ is applied.
} 
\label{figtwo}
\end{center}
\end{figure}

We take $A = S^+(j)$ 
for a single site $j$ in the center of a long chain. This operator 
constructs
a localized wavepacket consisting of all wavevectors. This packet spreads out
as time progresses, with different components moving at different speeds. 
The speed of a component is its group velocity, determined as the slope
of the dispersion curve at $k$.
In Fig.\ref{figtwo} we show the local magnetization 
$\bra\psi(t)|S^z|\psi(t)\ket$ for a chain of length $L=200$, with timestep $\tau=0.1$.
At $t=0$, the wavepacket has
a finite extent, with size given by the spin-spin correlation length $\xi$.
At later times, the different speeds of the different components give 
the irregular oscillations in the center of the packet.
We kept $m=150$ states per block, giving a truncation error of about 
$6\times10^{-6}$. 

>From this type of simulation we can construct the Green's function 
\begin{equation}
G(x,t) = -i \bra \phi| T[S^-_x(t) S^+_0(0)] |\phi \ket
\end{equation}
as $G(x,t) = -i \bra \phi(|t|)| S^-_x |\psi(|t|) \ket$.
Here $x$ is measured from the site $j$ where $S^+$ is applied.
We make one measurement of $G(x,t)$ for each left-to-right DMRG step, namely 
for step $x$. For efficiency we measure as we evolve in time, rather than,
say, devoting every other sweep to measurements without time evolving.
This may worsen the Suzuki-Trotter error somewhat, but we have found
the results quite satisfactory.
Since $G(x,t)$ is even in $x$ and $t$, the Fourier transform is
\begin{equation}
G(k,\omega) = 2 \int_0^\infty dt \cos \omega t \sum_x \cos kx G(x,t)
\end{equation}
The spectral function of interest is $-1/\pi {\rm Im} G(k,\omega)$.
We inevitably have some cutoff in time $T$ for the available data.
The maximum useable value of $T$ depends on the length of the chain: when the
leading edge of the wavefront hits the ends of the chain, the data 
no longer describes an infinite chain. On the other hand, before that
point the data does describe an infinite chain with boundary effects
dying off exponentially from the edges. This allows us to precisely
specify the momentum $k$, for times $t<T$.
To perform the time integral we multiply $G(x,t)$ by a 
windowing function $W(t)$
which goes smoothly to zero as $t \to T$. We have chosen a Gaussian,
$\exp(-4 (t/T)^2)$, which has the advantage of having a nonnegative
Fourier transform, yielding a nonnegative spectral function (except
for possible terms of size $\exp(-4)$). Note that
if the true spectral function has an isolated delta function peak, 
the windowed spectrum will have a Gaussian peak centered 
precisely at the same frequency.  Thus it is possible to locate the
single magnon line with an accuracy much better than $1/T$.
If a continuum is also present nearby, the peak is less well determined. 
In the case of the $S=1$ chain, for $k$ near $\pi$ the peak is isolated,
but at some point near $k=0.3 \pi$ the peak enters the two magnon continuum
and develops a finite width.  
Note that from our single simulation we
determine the spectral function for a continuum of values of $k$ and 
$\omega$.

\begin{figure}
\begin{center}
\epsfig{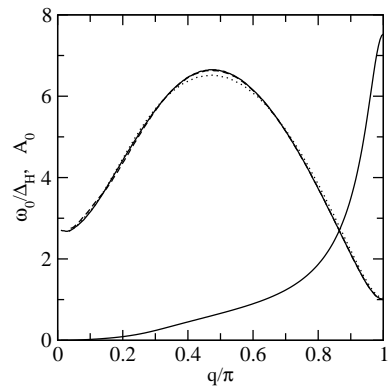}
\caption{
The single magnon line of the spin-1 Heisenberg antiferromagnetic chain.
The entire spectrum is obtained from one DMRG run, by Fourier transforming
the time and position dependent correlation function $\bra S^-_l(t) S^+_0(0)\ket$.
The broad solid curve shows the location of the maximum in the spectra for
a particular $q$, in units of the Haldane gap, 0.41050(2),
for a system of $L=600$ sites, using a time step $\tau=0.02$, running for $T=27.3$,
and keeping $m=200$ states. For comparison, results from two other runs are shown:
$L=400$, $\tau=0.1$, $T=60$, and $m=150$ (dashed curve); and.
$L=400$, $\tau=0.4$, $T=72$, and $m=200$ (dotted curve).
The solid curve peaked at $q=\pi$, shown only for the first run, 
is the weight $A_0$ in this quasiparticle peak, i.e.
$S(\omega) \approx A_0 \delta(\omega-\omega_0)$. 
} 
\label{figthree}
\end{center}
\end{figure}

In Fig.\ref{figthree} we summarize the results for the single magnon peak, 
determined automatically as the maximum of the spectrum.
To gauge the errors
we present results for several runs with various parameters. The results
are very close; the most signifcant errors are due to a finite $\tau$,
with the $\tau=0.4$ run showing some non-negligible errors. 
The results agree very well with accurate frequency-based
DMRG results\cite{kuehner} and quantum Monte Carlo\cite{montecarlo}.

\begin{figure}
\begin{center}
\epsfig{file=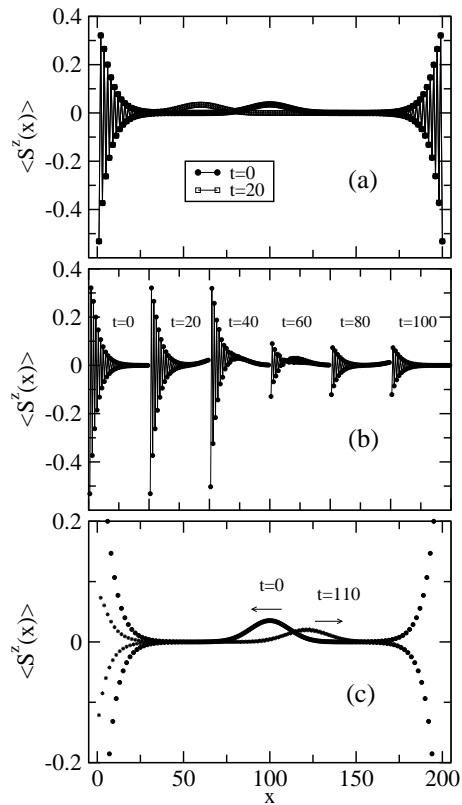,angle=0,width=6cm}
\caption{
A gaussian magnon wavepacket with momentum $k=0.8\pi$, $S^z=1$, and halfwidth 16
scattering off the left end of a
200 site spin-1 chain. The chain end states initially have $S^z=-1/2$.
In all cases we measure $\bra S^z(x) \ket$ at time $t$, and $\tau=0.2$.
In (a) we show $t=0$ and $t=20$. In (b), we show the left-most 50 sites for
a number of times. In (c), we show the whole chain for $t=0$ and $t=110$.
After the scattering, the system is in a non-local superposition of spin-flip and 
non-spin-flip states.
} 
\label{figfour}
\end{center}
\end{figure}

With real-time dynamics, we can simulate processes which would be very
difficult to understand via frequency dynamics. As an example, we
consider a magnon wavepacket scattering off the end of a spin-1 chain,
shown in Fig.\ref{figfour}. 
The magnon is a triplet, with $S^z=1$, and travels to the left with a 
speed of about 2.0.
The open ends of a spin-1 chain have spin-$1/2$ degrees of freedom, which
have received considerable attention \cite{aklt,huse}. One can view this end state
as a spinon bound to the end. An antiferromagnetic oscillation accompanies
this  state, decaying exponentially with the correlation length away 
from the edge. We choose the ground state with total spin $S^z=-1$, making the end 
states
each have $S^z=-1/2$. When the wavepacket hits the left end, it can scatter
either with or without a spin flip occuring. If the spin flip occurs, the
end spin changes to $S^z=1/2$ and the wavepacket to $S^z=0$. We see from
the figure that after the scattering, the end spin seems to have taken on 
an intermediate value of $S^z$, in particular $\bra S^z \ket \approx -0.11$.
Meanwhile, the wavepacket seems to have a total spin of 
$\bra S^z \ket \approx 0.61$.
Angular momenta of each are still quantized. The intermediate values occur because
we are observing a ``macroscopic'' quantum superposition of the state
with and without the spin flip. Specifically, the scattering is described by
\begin{equation}
|\rm{-}\frac{1}{2}, 1 \ket \to
a |\rm{-}\frac{1}{2}, 1 \ket  +
b |\frac{1}{2}, 0 \ket 
\end{equation}
where $a^2 \approx 0.61$, $b^2 \approx 0.39$. Note that the scattered $S^z=0$ magnon
does not show up when we measure $S^z_x$. A local description of
the wavefunction, as 
$(\alpha |\rm{-}\frac{1}{2}\ket + \beta |\frac{1}{2}\ket) \times
(\gamma |1 \ket + \delta |0\ket)$, is not possible; it does not conserve
total $S^z$. Our measurement of $S^z_x$,  $\bra \psi(t)| S^z_x |\psi(t) \ket$,
does not affect the state, but if one performed a real experiment and measured
the spin of, say, the magnon after scattering one would obtain either $S^z=1$ or
$S^z=0$. We find it quite remarkable that we can simulate a process in which 
such a non-local superposition develops! Our results raise the possibility of
using such spin chains for experimental studies of quantum measurement
and quantum computation.

We acknowledge the support of the NSF under grants
DMR-0311843.


\end{document}